\documentclass[aip,apl,reprint,longbibliography,amsmath]{revtex4-1}
\usepackage{graphicx}
\usepackage{todonotes}
\usepackage{upquote}

\begin{document}

\hyphenation{BASEX} 

\title{A direct comparison of high-speed methods for the numerical Abel transform}

\author{Daniel~D.~Hickstein} \email[]{danhickstein@gmail.com, Questions and comments regarding PyAbel may be posted to: \texttt{http://github.com/PyAbel/PyAbel} } \affiliation{Kapteyn--Murnane Laboratories Inc., Boulder, CO, U.S.A.}
\author{Stephen~T.~Gibson} \affiliation{Research School of Physics and Engineering, The Australian National University, Canberra 2601, Australia\looseness=-1}
\author{Roman Yurchak} \affiliation{Symerio, Palaiseau, France}
\author{Dhrubajyoti~D.~Das} \affiliation{Yale University, Dept. of Chemical and Environmental Engineering, New Haven, CT, U.S.A.\looseness=-1}
\author{Mikhail~Ryazanov} \affiliation{JILA, National Institute of Standards and Technology and University of Colorado, Boulder, CO 80309, U.S.A.\looseness=-1}


\date{\today}

\begin{abstract}
The Abel transform is a mathematical operation that transforms a cylindrically symmetric three-dimensional (3D) object into its two-dimensional (2D) projection. The inverse Abel transform reconstructs the 3D object from the 2D projection. Abel transforms have wide application across numerous fields of science, especially chemical physics, astronomy, and the study of laser-plasma plumes. Consequently, many numerical methods for the Abel transform have been developed, which makes it challenging to select the ideal method for a specific application. In this work eight transform methods have been incorporated into a single, open-source Python software package (PyAbel) to provide a direct comparison of the capabilities, advantages, and relative computational efficiency of each transform method. Most of the tested methods provide similar, high-quality results. However, the computational efficiency varies across several orders of magnitude. By optimizing the algorithms, we find that some transform methods are sufficiently fast to transform 1-megapixel images at more than 100~frames per second on a desktop personal computer. In addition, we demonstrate the transform of gigapixel images. 

\end{abstract}

\maketitle

\section{Introduction}
\begin{figure}
	\includegraphics[width=\linewidth]{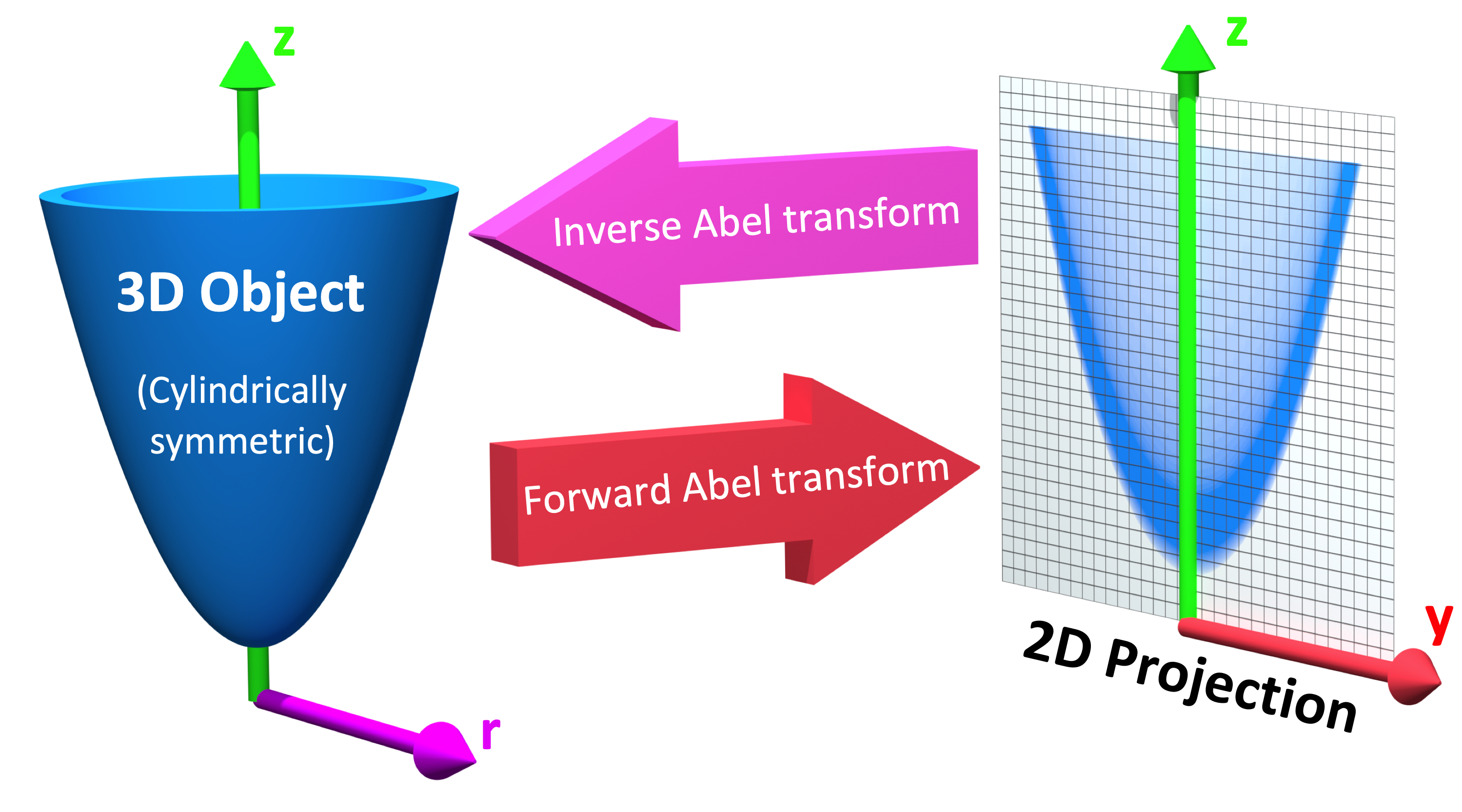}
	\caption{\label{fig:overview} The Abel transform maps a cylindrically symmetric three-dimensional (3D) object to its two-dimensional (2D) projection, a physical process that occurs in many experimental situations. For example, an x-ray image of the object on the left would produce the projection shown on the right. The \textit{inverse} Abel transform takes the 2D projection and mathematically reconstructs the 3D object. As indicated by Eqs.~(\ref{eq:forward}) and (\ref{eq:inverse}), the 3D object is described in terms of $(r,z)$ coordinates, while the 2D projection is recorded in $(y,z)$ coordinates.}
\end{figure}

The projection of a three-dimensional (3D) object onto a two-dimensional (2D) surface takes place in many measurement processes; a simple example is the recording of an X-ray image of a soup bowl, donut, egg, wineglass, or other cylindrically symmetric object (Fig.~\ref{fig:overview}), where the axis of cylindrical symmetry is parallel to the plane of the detector. Such a projection is an example of a \textit{forward} Abel transform and occurs in numerous experiments, including photoelectron/photoion spectroscopy \cite{dribinski2002, bordas1996, rallis2014, chandler1987, ryazanov2012, renth2006}, the analysis of plasma plumes \cite{glasser1978}, the analysis of flames \cite{deiluliis1998, cignoli2001, snelling1999, daun2006, liu2014, das2017}, and solar occultation of planetary atmospheres~\cite{gladstone2016, lumpe2007, craig1979}. The analysis of data from these experiments requires the use of the \textit{inverse} Abel transform to recover the 3D object from its 2D projection.

While the forward and inverse Abel transforms may be written as simple, analytical equations, attempts to naively evaluate these integrals numerically for experimental images does not yield reliable results \cite{whitaker2003}. Consequently, many numerical methods  have been developed to provide approximate solutions to the Abel transform \cite{dribinski2002, bordas1996,chandler1987,dasch1992,rallis2014,gerber2013,harrison2018,demicheli2017, dick2014}. Each method was created with specific goals in mind, with some taking advantage of pre-existing knowledge about the shape of the object, some prioritizing robustness to noise, and others offering enhanced computational efficiency. Unfortunately, each algorithm is implemented with somewhat different mathematical conventions and with often-conflicting requirements for the size and format of the input data. Additionally, the algorithms are written in different computer programming languages, and some only run on specific computing platforms. This situation makes it difficult to select the optimal Abel-transform method, since it can be very time-consuming to test multiple methods. Moreover, it is often unclear whether the observed differences between methods are intrinsic to the algorithm or simply a result of a particular software implementation.

In this work, we present the PyAbel package, which provides a consistent interface for the Abel-transform methods via the Python programming language. Within PyAbel, the transform methods share the same mathematical conventions and data format, which allows a straightforward, quantitative comparison of the output. In addition, this package is independent of the computing platform and has been tested on Linux, MacOS, and Windows. We find that, in general, the results of the various algorithms are similar, but that some methods produce inverse Abel transforms with somewhat less noise and in better agreement with analytical solutions.

In the process of implementing these transform methods in a common language, the numerical approaches were refined to optimize efficiency. After optimization, most of the transform methods operated more quickly than the available published results would suggest, often by several orders of magnitude. Previous studies \cite{rallis2014, harrison2018} have demonstrated ``real-time'' inverse Abel transforms of $1000\times1000$-pixel images at a rate of about 1 frame per second (fps), for a throughput of 1~megapixel per second (Mp/s). While this is sufficient for some applications, even inexpensive digital cameras can record and transfer high-definition video at frame rates of 30~fps or more, for a throughput of $\sim$50~Mp/s, and specialized high-speed cameras can exceed 1000~Mp/s. Consequently, there is a need for software that can perform the inverse Abel transform at high frame rates, preferably without requiring costly supercomputing infrastructure. Such an ability could be especially powerful when the results of an Abel transform are used in a feedback loop, for example, to optimize laser pulses to control a chemical reaction \cite{rallis2014}. 

This work demonstrates the first implementation of an inverse Abel transform operating at high-definition video rates, achieved using a standard desktop computer. In addition, the inverse Abel transform of a $65\,537\times65\,537$-pixel image is computed, pushing Abel transforms into the gigapixel range, for the first time. Additionally, a side-by-side comparison of the results of numerous methods for the inverse Abel transform is presented for both an analytical Gaussian function and experimental photoelectron-spectroscopy data.

\section{Abel-transform algorithms}
The forward Abel transform is given by
\begin{equation} \label{eq:forward}
F(y,z) = 2 \int_y^{\infty} \frac{f(r,z)\,r}{\sqrt{r^2-y^2}}\,dr,
\end{equation}
where $y$, $r$, and $z$ are the spatial coordinates as shown in Fig.~\ref{fig:overview}, $f(r,z)$ is the density of the 3D object at $(r,z)$, and $F(y,z)$ is the intensity of the projection in the 2D plane. The inverse Abel transform is given by
\begin{equation} \label{eq:inverse}
f(r,z) = -\frac{1}{\pi} \int_r^{\infty} \frac{dF(y,z)}{dy}\, \frac{1}{\sqrt{y^2-r^2}}\,dy.
\end{equation}

While the transform equations can be evaluated analytically for some mathematical functions, experiments typically generate discrete data (e.g., images collected with a digital camera), which must be evaluated numerically. Several issues arise when attempting to evaluate the Abel transform numerically. First, the simplest computational interpretation of Eq.~(\ref{eq:inverse}) involves three loops: over $z$, $r$, and $y$, respectively. Such nested loops can be computationally expensive. Additionally, $y = r$ presents a singularity where the denominator goes to zero and the integrand goes to infinity. Finally, a simple approach requires a large number of sampling points in order to provide an accurate transform. Indeed, a simple numerical integration of the above equations has been shown to provide unreliable results\cite{whitaker2003}.  

Various algorithms have been developed to address these issues. PyAbel incorporates eight algorithms for the inverse Abel transform, and three of these algorithms also support the forward Abel transform. Here we focus on the results of the inverse Abel transform, because it is the inverse Abel transform that is used most frequently to interpret experimental data. 

In the following, we describe the basic approach and characteristics of each transform method. The title of each method is the keyword for the method used in PyAbel.  Methods that pre-compute matrices for a specific image size -- and (optionally) save them to disk for subsequent reuse -- are indicated with an asterisk (*). All methods implement the inverse Abel transform, while methods that also implement a forward transform are indicated with a superscript F (\textsuperscript{\textbf{F}}).

\texttt{basex}*\textsuperscript{F} -- The ``BAsis Set EXpansion'' (BASEX) method of Dribinski and co-workers \cite{dribinski2002} uses a basis set of Gaussian-like functions. This is one of the \textit{de facto} standard methods in photoelectron/photoion spectroscopy \cite{whitaker2003}. The number of basis functions and their width can be varied. However, following the basis set provided with the original \texttt{BASEX.exe} program, by default we use a basis set where the full width at $1/e^2$ of the maximum is equal to 2~pixels and the basis functions are located at each pixel. Thus, the resolution of the image is roughly maintained. The \texttt{basex} algorithms allows a ``Tikhonov regularization'' to be applied, which suppresses intensity oscillations, producing a less noisy image \cite{dribinski2002}. In the comparisons presented here, the Tikhonov regularization factor is set to 200, which provides reasonable suppression of noise while still preserving the fine features in the image.

\texttt{hansenlaw}\textsuperscript{F} -- The recursive method of Hansen and Law \cite{hansen1985, hansen1985b, gascooke2000} interprets the Abel transform
as a linear space-variant state-variable equation, to
provide a reliable, computationally efficient transform. The \texttt{hansenlaw} method also provides an efficient forward Abel transform. 

\texttt{three\_point}* -- This ``three-point method'' and the following two methods (\texttt{two\_point} and \texttt{onion\_peeling}) are adapted from the 1992 paper by Dasch\cite{dasch1992}. All of these methods reduce the core Abel transform to a simple matrix-algebra operation, which allows a computationally efficient transform. Dasch emphasizes that these techniques work best in cases where the difference between adjacent points is much greater than the noise in the projections (i.e., where the raw data is not oversampled). The \texttt{three\_point} method provides a fast and robust transform by exploiting the observation that underlying radial distribution is primarily determined from changes in the line-of-sight projection data in the neighborhood of each radial data point. 

\texttt{two\_point}* -- The ``two-point method'' (also described by Dasch \cite{dasch1992}) is a simplified version of the \texttt{three\_point} algorithm and provides similar transform speeds.

\texttt{onion\_peeling}* -- Similar to the previous two methods, the ``onion-peeling deconvolution'' method described by Dasch \cite{dasch1992} is one of the simpler and faster inverse Abel-transform methods.

\texttt{onion\_bordas} -- The onion-peeling method of Bordas et al. \cite{bordas1996} is based on the MatLab implementation of Rallis et al. \cite{rallis2014}. While it is conceptually similar to \texttt{onion\_peeling}, the numerical implementation is significantly different.

\texttt{linbasex} -- The ``lin-BASEX'' method of Gerber et al.~\cite{gerber2013} models the 2D projection using spherical functions, which evolve slowly as a function of polar angle. Thus, it can offer a substantial increase in signal-to-noise ratio in many situations, but it is only appropriate for transforming projections that are appropriately described by the basis functions. This is the case for typical velocity-map-imaging photoelectron/photoion spectroscopy \cite{chandler1987} experiments, for which the algorithm was designed. However, for example, it would not be appropriate for transforming the object shown in Fig.~\ref{fig:overview}. The algorithm directly produces the coefficients of the involved spherical functions, which allows both the angular and radially integrated distributions to be produced analytically. This ability, combined with the strong noise-suppressing capability of using smooth basis functions, aids the interpretation of photoelectron/photoion distributions.

\texttt{direct}\textsuperscript{F} -- The ``direct'' methods \cite{yurchak2015} use a simple numerical integration, which closely resembles the basic Abel-transform equation (Eqs.~(\ref{eq:forward}) and (\ref{eq:inverse})). If the \texttt{direct} method is used in its most naive form, the agreement with analytical solutions is poor, due to the singularity in the integral when $r=y$. However, a correction can be applied, where the function is assumed to be piecewise-linear across the pixel where this condition is met. This simple approximation allows a reasonably accurate transform to be completed. Fundamentally, the \texttt{direct} method requires that the input function be finely sampled to achieve good results. PyAbel incorporates two implementations of the \texttt{direct} algorithm, which produce identical results, but with different calculation speeds. The \texttt{direct\_Python} implementation is written in pure Python, for easy interpretation and modification. The \texttt{direct\_C} implementation is written in Cython \cite{behnel2011}, a Python-like language that is converted to C and compiled, providing higher computational efficiency.

\section{Implementation}
\subsection{Interface}
PyAbel incorporates a streamlined interface to all of the transform methods, as well as numerous related functions for centering, symmetrizing, and circularizing the input images. Tools for analyzing the reconstructed images, including functions for angular and radial integration are also included. The ability to provide identical image preparation and output processing allows a quantitative comparison of transform methods. 

Generating a sample image, performing a forward Abel transform, and completing an inverse Abel transform requires just a few lines of Python code:


\begin{verbatim}
import abel
im0 = abel.tools.analytical.SampleImage().image
im1 = abel.Transform(im0, 
           direction ='forward', 
           method ='hansenlaw').transform
im2 = abel.Transform(im1,
           direction='inverse',
           method='three_point').transform
\end{verbatim}

Choosing a different method for the forward or inverse transform requires only that the \texttt{method} argument be changed. Additional arguments can be passed to the individual transform functions using the \texttt{transform\_options} keyword. A basic graphical user interface (GUI) for PyAbel is also available in \texttt{examples} directory.

In addition to the transform methods themselves, PyAbel provides many of the pre-processing methods required to obtain optimal Abel transforms. For example, an accurate Abel transform requires that the center of the image is properly identified. Several approaches allow to perform this identification in PyAbel, including the center-of-mass, convolution, and Gaussian-fitting. Additionally, PyAbel incorporates a ``circularization'' method, in the style of that presented by Gascooke \emph{et al.} \cite{gascooke2017}, which allows the correction of images that contain features that are expected to be circular (such as photoelectron and photoion momentum distributions). Moreover, the \texttt{pyabel.tools} module contains a host of \textit{post}-processing algorithms, which provide, for example, efficient projection into polar coordinates and radial or angular integration. A detailed project documentation can be found at \url{https://pyabel.readthedocs.io}.

\subsection{Conventions}

In order to provide similar results, we have ensured that the numerical conventions are consistent across the various transform methods. When dealing with pixel data, an ambiguity arises: do intensity values of the pixels represent the value of the data at $r=(0,\,1,\,2,\,...,\,n-1)$, where $n$ is an integer, or do they correspond to $r=(0.5,\, 1.5,\, 2.5, \,..., \,n-0.5)$? Either convention is reasonable, but comparing results from methods that adopt differing conventions can lead to small but significant shifts. We adopt the convention that the pixel values correspond to $r=(0,\,1,\,2,\,...,\,n-1)$. One consequence of this is that, when considering an experimental image that contains both the left and right sides of the image, the total image width must be odd, such that $r=(1-n, \, ..., \, -2, \, -1,\, 0,\,1,\,2,\,...,\,n-1)$. A potential disadvantage of our ``odd image'' convention is that 2D detectors typically have a grid of pixels with an \textit{even} width (for example, a $512\times512$-pixel camera). If the image were perfectly centered on the detector, our convention would not match the data, and a half-pixel shift would be required. However, in most real-world experiments, the image is not perfectly centered on the detector and a shift of several pixels is required, so the additional half-pixel shift is of no significance. 

A similar ambiguity exists with regards to the left--right and top--bottom symmetry of the image. In principle, since the Abel transform assumes cylindrical symmetry, left--right symmetry should always exist, and it should only be necessary to record one side of the projection. However, many experiments record both sides of the projection. Additionally, many experiments record object that possess top--bottom symmetry. Thus, in some situations, it is best to average all of the image quadrants into a single quadrant and perform a single Abel transform on this quadrant. On the other hand, the quadrants may not be perfectly symmetric due to imperfections or noise in the experiment, and users may wish to Abel-transform each quadrant separately and select the quadrant that produces the highest-fidelity data. PyAbel offers full flexibility, providing the ability to enforce arbitrary top--bottom and left--right quadrant averaging, and to specify which quadrants are averaged. By default, each quadrant is processed separately and recombined into in composite image that does not assume either top--bottom or left--right symmetry. 

In our performance benchmarks, left--right symmetry is assumed, because this is the most common benchmark presented in other studies \cite{rallis2014, harrison2018}. However, the image size is listed as the width of a square image. For example, $n=513$ indicates the time for the transformation of a $513\times513$-pixel image with the axis of symmetry located in the center. Since the Abel transform makes the assumption of cylindrical symmetry, both sides of the image are identical, and it is sufficient to perform the Abel transform on only one side of the image, or on an average of the two sides. So, to complete an Abel transform of a typical $513\times513$-pixel image, it is only necessary to perform the Abel transform on a $513\times257$-pixel array.

Another fundamental question about real-world Abel transforms is whether negative values are allowed in the transform result. In most situations, negative values are not physical, and some implementations set all negative values to zero. PyAbel allows negative values, which keeps the transform methods linear and gives users the option to average images either before or after the Abel transform without causing a systematic error in the baseline. Suppression of negative values can easily be achieved by including \texttt{A[A<0]=0}.

\section{Comparison of transform methods}
Since numerous Abel-transform methods have been incorporated into the same interface, it is straightforward to directly compare the results. Consequently, a user could simply try all of the transform methods and see which produces the best results or performance for a specific application. Nevertheless, here we present a brief comparison of the various transform methods when they are applied to a simple Gaussian function and also when they are used to transform a high-resolution photoelectron-spectroscopy image.

\begin{figure}
	\includegraphics[width=\linewidth]{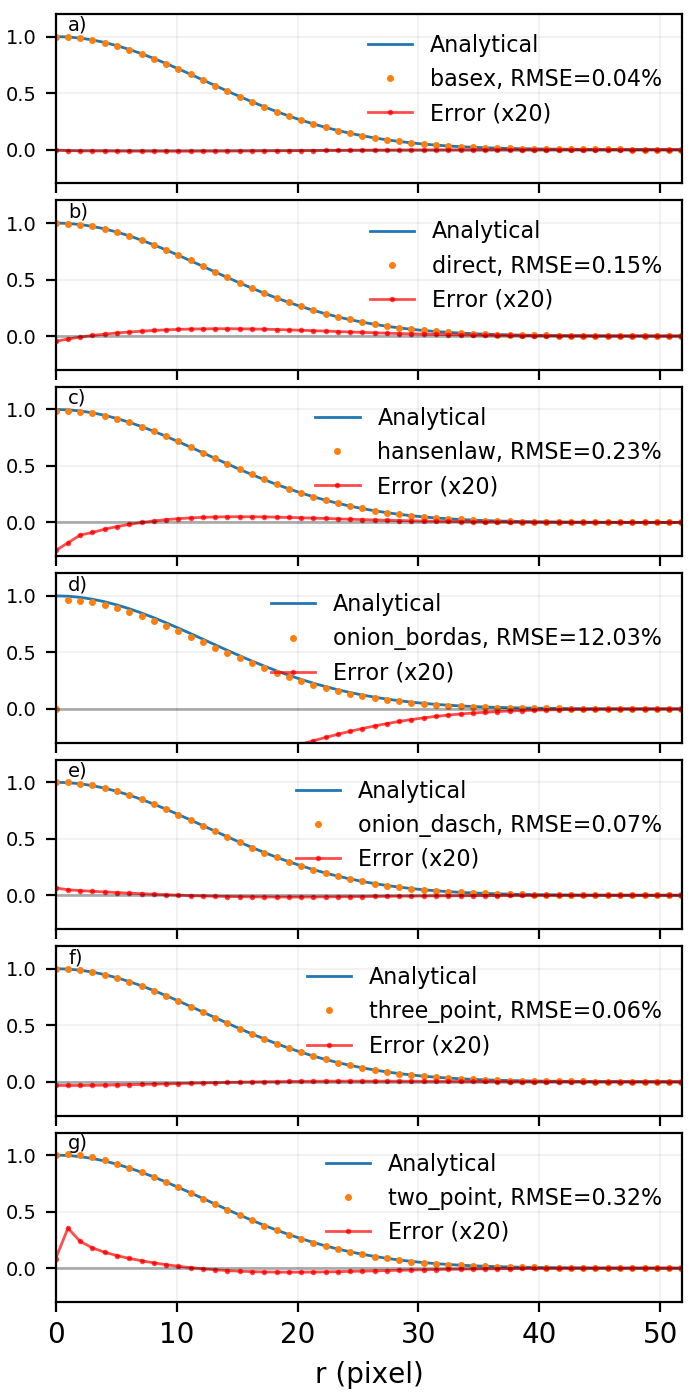}
	\caption{\label{fig:gaussian} \textbf{Comparison of inverse Abel-transform methods for a 1D Gaussian function with 70 points.} All of the inverse Abel transform methods show reasonable agreement for the inverse Abel transform of a Gaussian function. The root-mean-square error (RMSE) for each method is listed in the figure caption. In the limit of many pixels, the error goes to zero. However, when a small number of pixels is used, systematic errors are seen near the origin. This effect is more pronounced in some methods than others. The lowest error at the origin is seen from the method published in Dasch \cite{dasch1992} (\texttt{two\_point}, \texttt{three\_point}, and \texttt{onion\_peeling}) and the \texttt{direct} method. The \texttt{three\_point} method provides the lowest error. The \texttt{linbasex} method is not included in this figure because it is not applicable to 1D functions.}
\end{figure}

The Abel transform of a Gaussian is simply a Gaussian, which allows a comparison of each numerical transform method with the analytical result in the case of a one-dimensional (1D) Gaussian (Fig.~\ref{fig:gaussian}). As expected, each transform method exhibits a small discrepancy compared with the analytical result. However, as the number of pixels is increased, the agreement between the transform and the analytical result improves. Even with 70 points (the case shown in Fig.~\ref{fig:gaussian}), all of the method produce reasonable agreement, except at low $r$ values, where some methods diverge. While all methods show a systematic error as $r$ approaches zero, the \texttt{basex}, \texttt{three\_point}, and \texttt{onion\_dasch} seem to provide the best agreement with the analytical result. The direct methods show fairly good agreement with the analytical curve, which is a result of the ``correction'' discussed above. We note that the results from the \texttt{direct\_Python} and the \texttt{direct\_C} methods produce identical results to within a factor of $10^{-9}$.

\begin{figure}
	\includegraphics[width=\linewidth]{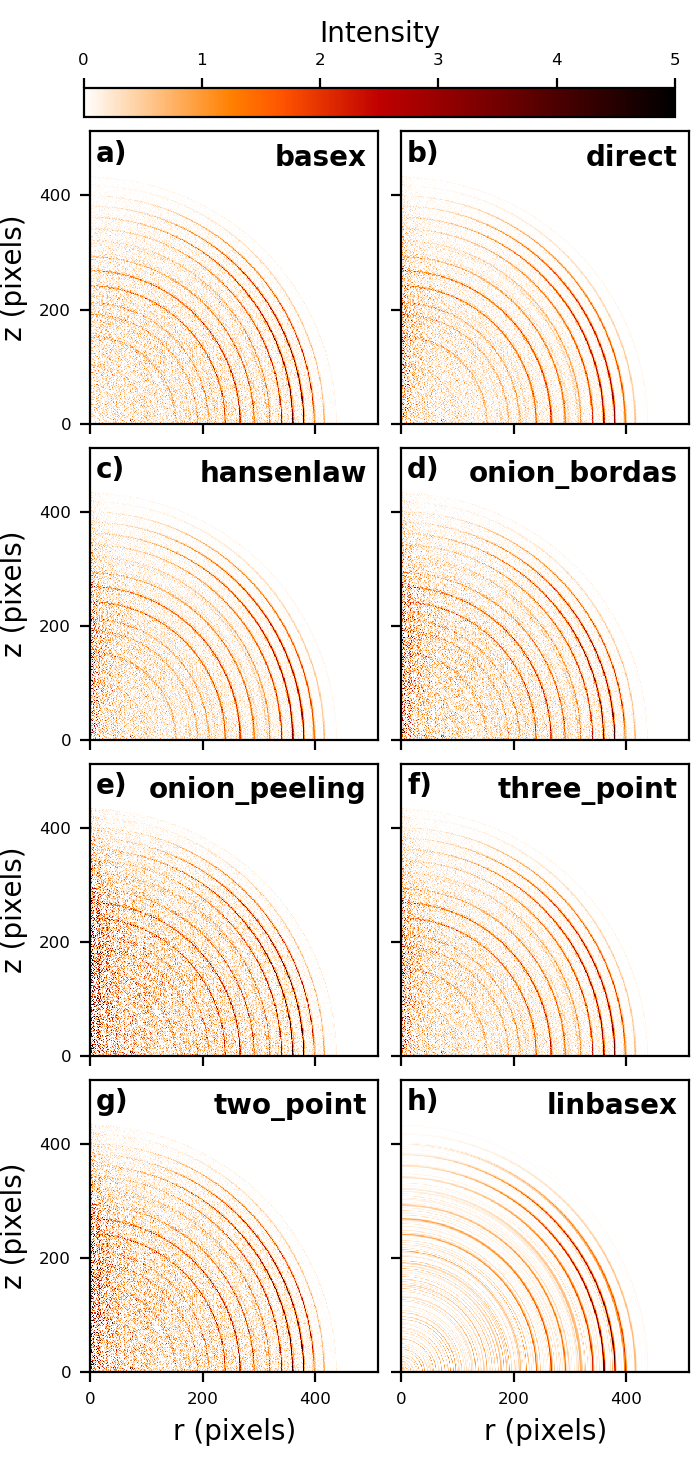}
	\caption{\label{fig:experiment} \textbf{Comparison of inverse Abel transform methods for an experimental photoelectron spectrum\cite{vanduzor2010}.} While all methods provide a faithful reconstruction of the original image, some of them cause a greater amplification of the noise present in the original image. The \texttt{linbasex} method models the image using a basis set of functions that vary slowly as a function of angle, which strongly reduces the high-frequency noise seen in the other transform methods. Besides the \texttt{basex} method with adjustable regularization, the \texttt{direct} and \texttt{three\_point} methods seem particularly suited for providing a low-noise transform. This dataset is the photoelectron spectrum of O$_2^-$ photodetachment using a 455~nm laser, as described in Ref~\onlinecite{vanduzor2010}.}
\end{figure}

Applying the various inverse Abel-transform methods to an experimental-photoelectron spectroscopy image (Fig.~\ref{fig:experiment}) provides a comparison of how the noise in the reconstructed image depends on the transform method. To a first approximation, the results of all the transform methods look similar. The \texttt{linbasex} method produces the ``smoothest'' image, which is a result of the fact that it models the projection using functions fitted to the image, that vary only slowly as a function of angle. The \texttt{basex} method incorporates a user-adjustable Tikhonov regularization factor, which tends to suppress noise, especially noise near the center of the image. Here, we set the regularization factor to 200, which provides significant noise suppression while providing no noticeable broadening of the narrow features. When the regularization factor is set to zero, the \texttt{basex} method provides a transform that appears very similar to the \texttt{onion\_peeling} method. For the other transform methods, the \texttt{direct} and \texttt{three\_point} methods appear to have the strongest noise-filtering properties. 

\begin{figure}
	\includegraphics[width=\linewidth]{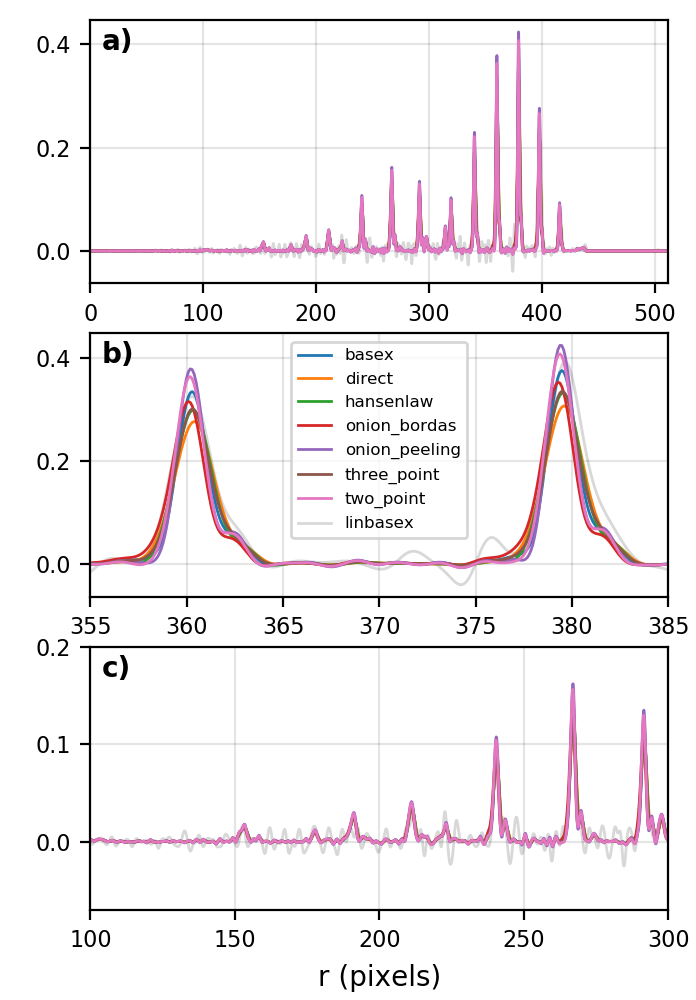}
	\caption{\label{fig:integration} \textbf{Comparison of inverse Abel-transform methods applied to an experimental photoelectron spectrum and angularly integrated.} The results shown in this figure are simply the angularly integrated 2D spectra shown Fig.~\ref{fig:experiment}. a)~Looking at the entire photoelectron momentum distribution, all of the transform methods appear to produce similar results. b)~Closely examining two of the peaks shows that most of the methods produce virtually identical results. c)~Examining the baseline across several peaks shows that most of the methods produce low amounts of noise between the peaks. }
\end{figure}

Fig.~\ref{fig:integration} shows the same data as Fig.~\ref{fig:experiment}, but with an angular integration performed to show the 1D photoelectron spectrum. Good agreement is seen between most of the methods, even on a one-pixel level. Interestingly, the \texttt{linbasex} method shows slightly more baseline noise than the other methods.

\section{Efficiency optimization}
\label{sec:efficiency}
\subsection{High-level efficiency optimization}
For many applications of the inverse Abel transform, the speed at which transform can be completed is important. Even for users who are only aiming to transform a few images, the ability to perform Abel transforms efficiently may enable more effective data analysis. For example, users may want to explore many different schemes for noise removal, smoothing, centering, and circularization, and faster Abel-transform algorithms allow this parameter space to be explored more rapidly and effectively.

While PyAbel offers improvements to the raw computational efficiency of each transform method, it also provides improvements to the efficiency of the overall workflow, which are likely to provide a significant improvements for most applications. For example, since PyAbel provides a straightforward interface to switch between different transform methods, a comparison of the results from each method can easily be made and the fastest method that produces acceptable results can be selected. Additionally, PyAbel provides fast algorithms for angular and radial integration, which can be the rate-limiting step for some data-processing workflows.

In addition, when the computational efficiency of the various Abel transform methods is evaluated, a distinction must be made between those methods that benefit from pre-computing information for a specific image size (\texttt{basex}, \texttt{three\_point}, \texttt{two\_point}, \texttt{onion\_peeling}) and those that do not (\texttt{hansenlaw}, \texttt{direct}, \texttt{linbasex}). Often, the time required for the pre-computation is orders of magnitude longer than the time required to complete the transform. One solution to this problem is to pre-compute information for a specific image size and provide this data as part of the software. Indeed, the popular BASEX application includes a ``basis set'' for transforming $1000\times1000$-pixel images. While this approach relieves the end user of the computational cost of generating basis sets, it often means that the ideal basis set for efficiently transforming an image of a specific size is not available. Thus, padding is necessary for smaller images, resulting in increased computational time, while higher-resolution images must be downsampled or cropped. PyAbel provides the ability to pre-compute information for any image size and cache it to disk for future use. Moreover, a cached basis set intended for transforming a larger image can be automatically cropped for use on a smaller image, avoiding unnecessary computations. The \texttt{basex} algorithm in PyAbel also includes the ability to extend a basis set intended for transforming a smaller image for use on a larger image. This allows the ideal basis set to be efficiently generated for an arbitrary image size.

\subsection{Low-level computational efficiency}
Transforming very large images, or a large number of images, requires inverse Abel-transform methods with high computational efficiency. PyAbel is written in Python, a high-level programming language that is easy to read, understand, and modify. A common criticism of high-level interpreted (non-compiled) languages like Python is that they provide significantly lower computational efficiency than low-level compiled languages, such as C or Fortran. However, such slowdowns can be avoided by calling functions from optimized math libraries for the key operations that serve as bottlenecks. For most of the transform methods (and indeed, all of the fastest methods), PyAbel uses matrix algebra functions provided by the NumPy library \cite{vanderwalt2011}, which are, in turn, provided by the Basic Linear Algebra Subprograms (BLAS) library (for example, the open-source OpenBLAS, Intel's Math Kernel Library (MKL), or Apple's Accelerate Framework). Thus, the algorithms in PyAbel have comparable performance to optimized C/Fortran. One subtle consequence of this reliance on the BLAS algorithms is that the performance is dependent on the exact implementation of BLAS that is installed, and users seeking the highest level of performance may wish to experiment with different implementations. In our tests, the fastest benchmarks have been achieved with MKL and Accelerate Framework.

\begin{figure}
	\includegraphics[width=\linewidth]{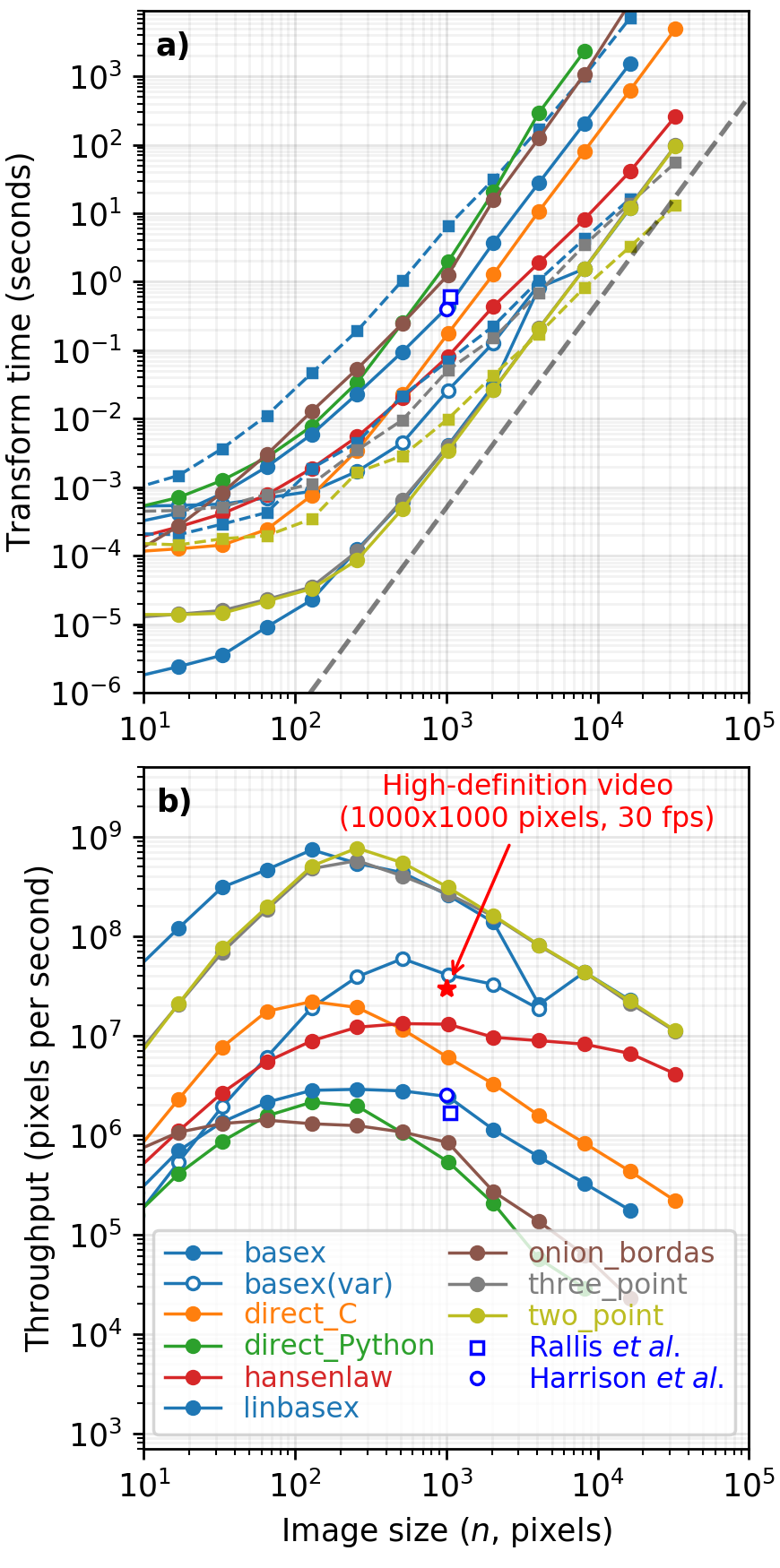}
	\caption{\label{fig:benchmark} \textbf{Computational efficiency of inverse Abel transform methods.} a) The time to complete an inverse Abel transform increases with the size of the image. The times for the generation of basis sets for the \texttt{basex} and \texttt{three\_point} methods are shown with dotted lines. Most of the methods display a roughly $n^{3}$ scaling (dashed gray line). b)~Alternatively, the performance can be viewed in terms of pixels-per-second rate. Here, it is clear that some methods provide sufficient throughput to transform images at high-definition video rates. These benchmarks were completed using a personal computer equipped with a 3.4~GHz Intel i7-6700 processor. For comparison, the transform times reported by Harrison \textit{et al.}\cite{harrison2018} (0.4~s for $n=1000$) and Rallis \textit{et al.} \cite{rallis2014} (0.6~s for $n\approx1054$) are shown as an open square/circle in panels a and b. }
\end{figure}

A systematic comparison of the time required to complete an inverse Abel transform versus the width of a square image is presented in Fig.~\ref{fig:benchmark}. We note that the \texttt{onion\_bordas} method completes a transform of a $1000\times1000$-pixel image in approximately 1 second, which is the same as reported by Rallis \emph{et al.} \cite{rallis2014}.

Fig.~\ref{fig:benchmark} reveals that the various methods display large differences in their computational efficiency. The \texttt{basex}, \texttt{two\_point}, \texttt{three\_point}, and \texttt{onion\_peeling} methods can run much faster if appropriately sized basis sets have previously been generated. The time for the basis-set generation for the \texttt{basex} and \texttt{three\_point} methods is also shown in the figure. The basis set generation for the \texttt{two\_point} and \texttt{onion\_peeling} methods take similar time to the \texttt{three\_point} method.

Fig.~\ref{fig:benchmark} shows the computational scaling as the image size is increased. A direct interpretation of the integral for the inverse Abel transform involves three nested loops, one over $z$, one over $r$, and one over $y$, and we should expect $n^3$ scaling. Indeed, the \texttt{direct\_C} and \texttt{direct\_Python} methods scale as nearly $n^3$. Several of the fastest methods (\texttt{onion\_peeling}, \texttt{two\_point}, and \texttt{three\_point}) rely on matrix dot-product operations. These methods scale roughly as $n^{3}$, which is approximately the expected scaling for matrix-multiplication methods \cite{coppersmith1990}. For typical image sizes ($\sim$500--1000 pixels width), \texttt{basex} and the methods of Dasch \cite{dasch1992} consistently out-perform other methods, often by several orders of magnitude. Interestingly, the \texttt{hansenlaw} algorithm exhibits a nearly $n^2$ scaling and should outperform other algorithms for large image sizes. 

Using a desktop computer (equipped with a 3.4~GHz Intel i7-6700 processor and running Linux) we are able to complete inverse Abel transforms with image sizes up to $32\,769\times32\,769$ pixels (Fig.~\ref{fig:benchmark}), for an image size of 1.07 gigapixels. Furthermore, using a dual-processor computer equipped with two 14-core 2.6~GHz Xeon E5-2697-v3 processors and 270~GB of memory, we were able to use both the \texttt{hansenlaw} and \texttt{three\_point} methods to complete transforms up $65\,537\times65\,537$ pixels, for an image size of 4.30 gigapixels. To our knowledge, this is the first demonstration of the numerical Abel transform of gigapixel-scale images. The capability of transforming large images is becoming increasingly important, as techniques for collecting gigapixel-scale images, such as mosaic imaging and large CCD arrays become more commonplace \cite{brady2012}.

\subsection{High-framerate transforms}
Many experiments that rely on inverse Abel transforms utilize ``high-definition'' video cameras that record data at a rate of more than 20~Mp/s. For example, a ``720p'' camera records $1280\times720$ images at 30 fps for a data-rate of 27.6~Mp/s. As shown in Fig.~\ref{fig:benchmark}b, the \texttt{basex}, \texttt{onion\_peelng}, \texttt{two\_point}, and \texttt{three\_point} methods are capable of performing an inverse Abel transform at data rates of nearly 100 Mp/s for $n\approx1000$, thus achieving Abel transforms at high-definition video rates for the first time.

Since the \texttt{basex} method has an adjustable regularization parameter, Fig.~\ref{fig:benchmark} also shows a ``basex(var)'' curve that corresponds to changing this parameter for each data frame. This requires additional computations and thus slows down the processing. Nevertheless, the throughput remains sufficient for high-definition video rates, meaning that the regularization parameter can be adjusted in real time. This capability is very helpful even for analyzing individual images, as users can vary the regularization strength and immediately observe how it affects the results.

\section{Conclusion}
Here we have presented the PyAbel software package for completing forward and inverse Abel transforms, which allows the numerical transformation of three-dimensional objects into their two-dimensional projection and \textit{vice versa}. We have implemented eight different algorithms for the inverse Abel transform and found good agreement between the results produces using the various methods. After significant optimization of the algorithms in each method, we analyzed the computational efficiency of each method and compared the scaling of the frame rate with the image size. We have made the first demonstration of inverse Abel transforms at high-definition video frame rates (1-megapixel images at more than 30 frames per second). In addition, we have transformed a $65\,537\times65\,537$-pixel image, realizing inverse Abel transforms of gigapixel images. Moreover, all of the Abel transform methods are implemented in Python (or Cython), a high-level scripting language that allows easy modification and incorporation into other projects. This ability to easily complete efficient Abel transforms should enable new capabilities in many fields.

PyAbel is open-source and freely available at \texttt{http://github.com/PyAbel/PyAbel}. The PyAbel development team encourages the incorporation of new Abel transform methods into PyAbel. 
\\

\begin{acknowledgments}
We acknowledge useful feedback from Eric Hansen, Jason Gascooke, Gilbert Shih, Eric Wells, Chris Rallis, Adi Natan, Kevin Dorney, Jennifer Ellis, and Quynh Nguyen.

S.T.G.'s research was supported by the Australian Research Council Discovery Project, Grant DP160102585.

\end{acknowledgments}

\section{References}
\bibliography{pyabel}

\end{document}